\documentclass[preprint]{aastex}

\usepackage{color}
\usepackage{natbib}
\usepackage{graphicx}
\usepackage{amssymb}
\usepackage{multirow}
\usepackage[varg]{txfonts}

\bibpunct{(}{)}{;}{a}{}{,}

\usepackage{hyperref}

\hypersetup{
    bookmarks=true,         
    unicode=false,          
    pdftoolbar=true,        
    pdfmenubar=true,        
    pdffitwindow=true,     
    pdfstartview={FitH},    
    pdftitle={},    
    pdfauthor={Beers},     
    pdfsubject={Astronomy},   
    pdfcreator={dvipdf},   
    pdfproducer={dvipdf}, 
    pdfkeywords={metal-poor stars},
    pdfnewwindow=true,      
    colorlinks=true,       
    linkcolor=red,          
    citecolor=blue,        
    filecolor=magenta,      
    urlcolor=cyan,           
    breaklinks=true,
    linktocpage
}

\begin{document}
\title{The Origin of the Milky Way's Halo Age Distribution} 

\author{Daniela Carollo}
\affil{Center of Exellence for All Sky Astrophysics (CAASTRO) - Australia\\}
\affil{INAF, Astrophysical Observatory of Turin, Torino, Italy\\ }
\email{danielle100596@gmail.com} 

\author{Patricia B. Tissera}
\affil{Departamento de Ciencias Fisicas, Universidad Andres Bello, 
Av. Republica 220, Santiago, Chile\\}
\affil{Millennium Institute of Astrophysics, Av. Republica 220, 
Santiago, Chile\\}

\author{Timothy C. Beers, Dmitrii Gudin}
\affil{Department of Physics and JINA Center for the Evolution of the 
Elements, University of Notre Dame, Notre Dame, IN  46556 USA\\
}

\author{ Brad K. Gibson}
\affil{E.~A Milne Centre for Astrophysics, University of Hull, Hull, 
HU6~7RX, United Kingdom\\}

\author{Ken C. Freeman}
\affil{ANU - Research School of Astronomy and Astrophysics, 
Weston - ACT, Australia\\}

\author{Antonela Monachesi}
\affil{Instituto de Investigaci\'on Multidisciplinario de Ciencia y Tecnolog\'ia, Universidad de La Serena, Ra\'ul Bitr\'an 1305, La Serena, Chile.\\}
\affil{Departamento de F\'isica y Astronom\'ia, Universidad de La Serena, Avda. Juan Cisternas 1200, La Serena, Chile}

\begin{abstract} 
We present an analysis of the radial age gradients for the stellar halos 
of five Milky Way mass-sized systems simulated as part of the Aquarius 
Project. The halos show a diversity of age trends, reflecting their 
different assembly histories. Four of the simulated halos possess clear 
negative age gradients, ranging from approximately $-$7 to 
$-$19~Myr/kpc , shallower than those determined 
by recent observational studies of the Milky Way's stellar halo. 
However, when restricting the analysis to the accreted component alone, 
all of the stellar halos exhibit a steeper negative age gradient with values 
ranging from $-$8 to $-$32~Myr/kpc, closer to those observed in the 
Galaxy. Two of the accretion-dominated simulated halos show a large 
concentration of old stars in the center, in agreement with the 
Ancient Chronographic Sphere reported observationally. The stellar halo 
that best reproduces the current observed characteristics of the age 
distributions of the Galaxy is that formed principally by the accretion 
of small satellite galaxies. Our findings suggest that the hierarchical clustering 
scenario can reproduce the MW's halo age distribution
if the stellar halo was assembled from accretion and disruption of satellite galaxies with dynamical masses
less than $\sim$10$^{9.5}$~M$_{\sun}$, and 
a minimal \it in situ \rm contribution.
\end{abstract}


\section{Introduction}

The stellar halos of galaxies play a particularly crucial role in
understanding the early formation of galaxies due to the dynamical and
chemical fingerprints they carry; fingerprints which shed light on a
given galaxy's assembly history and the chemical history of its accreted
satellite galaxies. In the Milky Way (MW), the advent of massive
photometric and spectroscopic surveys has revolutionized the
understanding of our own stellar halo. With respect to the diffuse halo, 
\citet{carollo2007,carollo2010}, and \citet{beers2012} demonstrated that 
it comprises at least two stellar components with different
kinematics, dynamics, and chemical composition  (the inner and
outer halos). These findings have now been supported by
a large number of authors \citep{dejong2010, deason2011, an2013,nissen2010, kafle2013,
hattori2013,an2015,das2016}, and most recently by 
\citet{helmi2016} using Gaia data. The MW's stellar halo is 
also populated by numerous stellar streams and overdensities, the
product of past mergers of satellite galaxies \citep{helmi2008,
bell2008}, suggesting a complex superposition of stellar populations.

Another crucial parameter that can provide a comprehensive picture of
the MW assembly process is the age distribution of the halo system.
Recently, \citet{santucci2015} determined the age of the underlying
stellar populations in the MW's halo out to $\sim$25~kpc from the
 Galactic center by employing a small sample ($\sim$4700) of
spectroscopically-confirmed blue horizontal-branch (BHB) stars
selected from the Sloan Digital Sky Survey (SDSS - DR8;
\citet{aihara2011}). The resulting age map showed that in the
central region of the MW, there exists a region (out to 10-15~kpc)
 composed of a significant concentration of very old stars
($\sim$11.5-12.5~Gyr), with younger structures or overdensities
extending out to greater distances (likely associated with the
Sagittarius stream and Virgo overdensity). In a later analysis, 
\citet{carollo2016} produced a high-resolution age map extending out to 
$\sim$60~kpc, using a much larger sample ($\sim$130,000) of
photometrically-selected BHB stars from SDSS-DR8. In addition to the
aforementioned central region dominated by old stars, an inferred age
gradient of $-$25$\pm$1~Myr~kpc$^{-1}$ was derived, and numerous known
and unknown structures and overdensities were identified. The existence and claimed value of a
negative age gradient was confirmed by \citet{das2016} in an
investigation based on a sample of spectroscopically-identified
BHB stars from SDSS-DR8.
 
In cosmological simulations of MW-mass galaxies, it is possible to
follow the evolution of the satellite galaxies
 which merge as part
of a system's assembly \citep{brook2004,bullock2005,cooper2010,
few2012,few2014,gomez2012,tissera2014,cooper2015}. In this scenario,
 stellar halos are predicted to form primarily through the
accretion of satellite galaxies, each with different stellar
masses, gas fractions, and stellar population distributions
\citep[e.g.,][]{tissera2014, cooper2015}. An important contribution
from \it in situ \rm stars within the inner $\sim $15 kpc has been also
identified in hydrodynamical simulations, which have a range of possible
origins 
\citep[e.g.,][]{zolotov2009,font2011,house2011,brook2012,tissera2013,cooper2015,
pillepich2015,monachesi2016b}. 


Recent simulations of MW-mass galaxies provide information on the age
distribution of stars formed both \it in situ \rm and in accreted
satellites, however, most such studies do not include a discussion of
the overall age gradients \citep[e.g.,][]{brook2012,tissera2013,
pillepich2015}. This was in part driven by the lack of any strong empirical
constraints with which to compare. One early exception is the
semi-cosmological, sticky-particle work of \citet{bekki2001}, who found
an age gradient of $-$30~Myr~kpc$^{-1}$ over the galactocentric radial
range of 20-50~kpc in one low-resolution simulation \citep[see
also][]{samland2004}.

In this paper, we focus on the age structure of the stellar halo system 
in a set of five simulated MW mass-sized halos from the Aquarius 
Project. Understanding the age structure of the stellar halos can 
contribute to reveal their origin and evolution. These stellar halos have 
been considered extensively in previous papers, including their chemical 
abundance patterns, density and metallicity profiles, and assembly 
histories \citep{tissera2012,tissera2013,tissera2014,tissera2016}. Here, 
the age gradients and the relative contributions of the \it in situ \rm 
and accreted components, as a function of the galactocentric radius, are 
explored. The aim of this analysis is to examine to what extent the 
simulated Aquarius halos are able to reproduce the age trends found in 
the MW's halo, and to investigate the connection between the age profiles 
and halo assembly. For these purposes, we select a subset of the 
level-5 runs from \citet{scan2009}, specifically Aq-A, Aq-B, Aq-C, Aq-D, 
and Aq-G. Although none of the five realizations underwent a recent 
major merger, this simulated set of galaxies produce an excess of stars, 
and consequently, more massive stellar halos than the MW. Such issues 
affect other simulations as well (e.g., \citet{font2011,
pillepich2015}), and should be addressed in future work, given that the MW is an ${\rm 
L^*}$ spiral galaxy with perhaps the least massive stellar halo for its 
given total mass \citep{harmsen2017}. 
 A more efficient supernova feedback can contribute to decrease the
  stellar mass fraction, and form more
  extended disk structures \citep[e.g.][]{aumer2013,pedrosa2015}.  We note, however, that
  the measured mass of the MW stellar halo may be underestimated by
  current observational methods (\citet{sanderson2017}). The
  analysis presented in this paper apply a new
  observational constraint; the MW
  stellar age distribution, a critical piece of information for the
  confrontation of models and observations, and contribute to
  understanding the origin of the MW's stellar halo. 
 This paper is organized as follows: Section~2 provides a short 
description of the simulations, while Section~3 describes the analysis 
of the age distributions and associated age maps. The summary and 
conclusions are reported in Section~4.

\section{The Simulated Aquarius Galaxies}

We analyzed a subset of five MW-mass galaxies from the Aquarius Project 
run with a version of {\small GADGET-3}, an optimized version of {\small GADGET-2} \citep{springel2005}, which was modified to include 
supernova feedback and chemical evolution by \citet{scan2005} and 
\citet{scan2006}. This code allows the description of a multiphase 
medium and the triggering of mass-loaded galactic outflows without 
introducing mass-dependent scaling parameters. The chemical model 
describes the enrichment by Type II (SNII) and Type Ia (SNIa) supernovae 
according to the nucleosynthesis yields of \citet{WW1995} and 
\citet{thie1993}, respectively. These simulations are explained in 
detail by \citet{scan2009}, and have been used throughout our ongoing 
series of papers 
\citep{scan2009,scan2010,tissera2012,tissera2013,tissera2014,tissera2018}; 
here, we only provide their main characteristics. The initial conditions 
are consistent with a $\Lambda$-CDM cosmology having the following 
parameters: $\Omega_{\rm m}=0.25,\Omega_{\Lambda}=0.75, \Omega_{\rm 
b}=0.04, \sigma_{8}=0.9, n_{s}=1$ and $H_0 = 100 \ h \ { \rm km~s^{-1} 
Mpc^{-1}}$, with $h =0.73$. The dark matter particle mass was 
10$^{6}$~${\rm M_\odot }$~$h^{-1}$ and the initial gas particle mass 
2$\times$10$^{5}$~${\rm M_\odot}$~$h^{-1}$. The corresponding 
gravitational softenings ranged from 0.5$-$1~kpc$~h^{-1}$. The halos 
were selected not to have had a major merger since $z$$<$2.
 
We aligned the disks with the xy plane as described  in \citet{tissera2012} and employed their adopted dynamical decomposition methodology.
We measured $\epsilon= J_{\rm z}$/J$_{\rm 
z,max}(E)$ for each star, where J$_{\rm z}$ is the angular momentum 
component perpendicular to the disk plane and J$_{z,max}(E)$ is the 
maximum J$_{z}$ over all particles of given total energy, E.  A star on 
a prograde circular orbit in the disk plane has $\epsilon =1$; stars 
with $\epsilon > 0.65$ are considered a part of the disk components. 
Particles which do not satisfy this requirement are taken to be part of 
the spheroidal components. Motivated by observations of the MW spheroid 
 that exhibit differences in stellar kinematics and chemical abundances as 
one moves outwards \citep{carollo2007,carollo2010,zoccali2008}, we 
separate our spheroids into two stellar populations according to their binding 
energy. The central spheroid (bulge) is defined by stars more 
bound than those with the minimum energy (E$_{cen}$) at r $\geq$ 0.5 x 
r$_{\rm opt}$ (r$_{\rm opt}$ is
  defined as the radius that encloses $\sim$80\% of the stellar mass
  identified by the SUBFIND algorithm \citep{springel2001}. 
 Stars more weakly bound than E$_{\rm cen}$ are taken as 
part of the stellar halo. In this paper, and to confront with 
observations, the stellar halos are not separated into inner and outer 
components, as done in previous works but are taken as an ensemble 
system. These criteria are chosen so that the definition of the 
 stellar halos adapts to the overall size of each individual 
galaxy, and is the same definition used in the aforementioned series of 
Aquarius Project stellar-halo papers. For the analysis that follows, we 
have removed satellite galaxies identified by the SUBFIND 
  as individual systems, but disrupted satellites and/or any residual stellar streams remain.

\citet{tissera2013} and \citet{tissera2014} analyzed the spatial 
distributions, chemical abundances, and formation histories of the 
stellar populations of the Aquarius halos and found evidence that they 
are mainly built by stars formed in satellite galaxies, and later accreted 
onto the main halo. However, and as in agreement with previous numerical 
results, an important contribution of \it in situ \rm stars are detected 
in the central regions (recall, \S1). In addition, \citet{tissera2014} 
showed that stellar halos 
dominated by accreted massive satellite galaxies 
(M$>$10$^{9.5}$~M$_{\odot}$) exhibit steeper metallicity gradients
 \citep[see also][]{cooper2010}.  
They also reported that low-metallicity stars are mainly contributed by 
low-mass satellites, and are more frequent in the outskirts of 
halos. The central regions of such systems (within $\sim$10~kpc, 
including the bulge) have been analyzed in detail by 
\citet{tissera2018}, where a significant contribution of old stars was 
found. Moreover, one of the simulated halos (Aq-C-5) showed a good 
match with the spatial, kinematic, and metallicity properties observed 
in the MW \citep{zoccali2016}. The analysis of the 
assembly history of the central regions shows that this halo did not 
accrete satellite galaxies more massive than $\sim$10$^{10}$~M$_\odot$ 
 during its assembly. This characteristic is also relevant in the 
analysis of the age gradients in our simulated halos.

\section{Analysis of the Age Distributions}

For the five simulated halos, we identify those stars that formed
\it in situ \rm or in accreted satellite galaxies by adopting the
following criteria: \it in situ \rm stars were assumed to form within
the virial radius, while accreted stars were assumed to form in separate
galaxies prior to accretion (i.e., before entering the virial
  radius of the progenitor galaxy). \citet{tissera2013} defined three different
sub-populations of \it in situ \rm stars: 1) those born from gas
accreted in the first stages of assembly, 2) disk-heated stars formed in
the disk structure of the main progenitor galaxy, then heated
kinematically, and 3) those formed from gas carried in by gas-rich
satellite galaxies (endo-debris).
 As
described in previous works, disk-heated and endo-debris stars exhibit distinct
chemical properties that can help to link observations to the
galaxy-formation models \citep{brook2012, tissera2013}. Nevertheless, in this paper, for the sake of
clarity and simplicity, all stars born inside the virial radius are
grouped and analyzed as \it in situ \rm stars, as they dominate
the inner region of the stellar halos, while the outer regions are
mainly populated by accreted stars\footnote{We checked that, if
  endo-debris stars were considered part of the accreted subsample,
  the age distributions would not change significantly.}.

 For illustration purposes, Figure 1 shows the smoothed
  stellar age-map distribution projected onto (x,z) plane (z
  is the direction of rotation of the galaxy) for three of the analyzed halos\footnote{The stellar age-maps are mass-weighted and smoothed using a spline kernel consistently with the hydrodynamical code used to perform the simulations.}. 
The maps have been built by removing the contribution of the bulge
 according to the criteria given in Section 2. The upper panels represent the entire population of 
stars assigned to the halos, while the lower panels show the maps for 
the accreted stellar population only; both maps extend to a radius of 
100~kpc. As can be seen, the halos exhibit clumpy age structure at large 
distances from the center, reflecting the mixture of stars with 
different ages. The presence of younger structures and their increasing 
number with galactocentric distance are globally consistent with the 
observational results of the MW reported in \citet{carollo2016}.  By 
comparing the age distributions of the entire stellar halo population (upper 
panels) with those of the accreted stars only (lower panels), different 
features emerge: when the \it in situ \rm stars are removed from the age 
map, an age distribution qualitatively more similar to that observed in the MW's halo is 
 visible, consisting of younger structures at large distances
 from the galactic center and a concentration of older stars in the 
central region. 

Examination of the panels in Figure 1 reveals that the accreted central 
region of Aq-C and Aq-A is dominated by old stellar populations (dark 
blue area), in agreement with the Ancient Chronographic Sphere (ACS)
observed in the MW by \citet{santucci2015} and
\citet{carollo2016}. We note that the {\it in situ} components also
  contribute with old stars as shown in \citet{tissera2018}.  
Apart from this common feature, each halo has its own peculiarities. In 
the case of Aq-C, when the entire stellar-halo population is considered, the 
central region of the galaxy exhibits a disk-shaped feature, due to the 
contribution of younger, disk-heated stars. In fact, the mass fraction of 
halo stars which originated in the disk is $\sim$40\%, 
consistent with the fractions found by \citet{brook2012}. Interestingly, 
Aq-D has a different age distribution, with slightly younger accreted 
stars, compared to the other two halos. 

To quantify the age variations, we estimate the age profiles for each
stellar component as a 
function of galactocentric radius. 
 The age profiles are derived by taking the median 
values in concentric shells  of 2~kpc radial extension.  Figure 2 shows the age trends for the \it in 
situ \rm (left panel), accreted (middle panel), and combined (right 
panel) stellar halo populations (all represented by solid lines). The age-gradient found in observational studies of the MW is denoted by the green dot-dashed line.
 As can be seen, the \it in situ \rm 
sub-populations possess, on average, flat age gradients, and in some cases 
(Aq-A, Aq-C, Aq-D) a significant contribution of younger stars in the very central 
regions (10-20~kpc). 
 Such contributions are caused by gas brought in by 
gas-rich satellite galaxies that enter the virial radius (these stars are 
 classified as the endo-debris population and included in the \it 
 in situ \rm component) or by disk-heated stars. Each age profile
 reflects its particular history of assembly.

The accreted stellar population exhibits a steeper age gradient in most 
of the analyzed halos, as can be seen from the linear regressions 
applied to the age profiles of these components (Fig. 2,
middle panels). In fact, the inspection of Fig.~2 reveals 
that the accreted components are younger than the \it in situ \rm 
stellar populations for Aq-B and Aq-G by 1$-$3~Gyr, while for Aq-A and 
Aq-C, the accreted component is younger than the \it in situ \rm one 
only in the outer halo region (r$>$25~kpc), by less than 1~Gyr.
 The accreted population in Aq-B exhibits a quadratic-shape age
gradient, but it does not appear statistically significant, while
model Aq-C shows a significant linear age gradient in the range from $\sim -11$ to $\sim -12$ Gyr (not labeled in the panel).

In the right column of Fig.~2, the total age profiles for the entire 
 stellar halo are shown together with the linear regression fits (solid black 
and red lines, respectively). Negative age gradients are found for Aq-A, 
Aq-B, and Aq-C, with values of $-$8.0 $\pm$ 2.0 Myr~kpc$^{-1}$, $-$12.3 
$\pm$ 3.5 Myr~kpc$^{-1}$, and $-$11.8 $\pm$ 1.0 Myr~kpc$^{-1}$ (not labeled in the panels), 
respectively, while Aq-D and Aq-G show very weak age gradients. The 
gradients derived for the {entire} stellar halos of the Aquarius set are 
shallower than those determined in observational studies of the MW's 
stellar halo\footnote{Note that in the MW's stellar halo, spherically 
averaged profiles cannot be estimated.}: $-$25~Myr~kpc$^{-1}$ 
\citep{carollo2016} and $-$30~Myr~kpc$^{-1}$ 
\citep{das2016}.\footnote{Errors on the age gradients are 1-2 Myr 
kpc$^{-1}$.}

It is worth noting that observations of halo stars, both within the MW 
and in external galaxies, are optimized to reduce disk-component 
contamination. In particular, observations carried out in external 
galaxies are generally performed along the major axis of disk rotation 
\citep{monachesi2016a,harmsen2017}. To better match the observational 
conditions, we re-calculate the age profiles and fractions by excluding 
all stars within 5~kpc of the mid-plane. By doing this, the influence of 
stars which might still belong to an extended vertical disk is minimized 
\citep{harmsen2017}. The age profiles for these sub-samples {(hereafter referred to as the
fiducial stellar halos)} are 
represented in Figure 2 with dot-dashed lines. As can be seen from the 
figure, most of the discarded stars belong to the \it in situ \rm 
stellar populations (larger discrepancies in the \it in situ \rm age 
profiles with respect to the original sample). The recalculated age 
profiles are, in most cases, steeper than those derived by adopting the 
entire halo sample 
and are labeled in the panels. The  age gradients for the fiducial
  stellar 
halo populations (\it in situ \rm and accreted stars combined) are, $-$6.8 $\pm$ 2.8 
Myr~kpc$^{-1}$, $-$19.5 $\pm$ 4.0 Myr~kpc$^{-1}$, $-$19.2 $\pm$ 1.3 
Myr~kpc$^{-1}$, $+$0.2 $\pm$ 1.0 Myr~kpc$^{-1}$, and $-$9.6 $\pm$ 1.3 
Myr~kpc$^{-1}$, for Aq-A, Aq-B, Aq-C, Aq-D, and Aq-G, respectively. 
These slopes are shaped significantly by the particular assembly history 
of each halo.

The global age trend in a given halo is affected not only by the median
age of the \it in situ \rm and accreted stars, but also by their
relative contribution as a function of the radius. Figure 3 shows the
stellar mass fraction vs. galactocentric radius for the \it in situ \rm
(red), accreted (green), and the total (black) stellar populations in
the simulated halos. The stellar mass fraction is calculated within the
virial radius for both the entire (solid line) and the
  fiducial (dot-dashed line) stellar halos. It is also important to mention that Aquarius stellar halos have a
contribution of stars younger than 10 Gyr representing $\sim
1\%-8\%$ of the total stellar halo mass,
except for Aq-B (29\%) which has experienced a more recent massive accretion. 
These stars are mostly associated
to the in situ component and to the accretion of more massive satellites. Such younger stellar population is not represented in current observations that make use of BHB stars to derive the age structure of the halo system. Nevertheless, in the
Aquarius simulations the presence of stars younger than 10 Gyr do not strongly affect the overall trends of the age gradients, except in the central regions.

As can be seen in Figure 3, the stellar mass fraction of the \it in situ 
\rm population is dominant out to $\sim$20~kpc for all the halos except 
Aq-B. Beyond $\sim$20~kpc, the accreted component dominates over the \it
in situ \rm one at all galactocentric distances, within the virial
radius, and for all the simulated halos. This is consistent with
the findings of previous works \citep[e.g.][]{tissera2014,cooper2015}. The
negative age trends are determined principally by the accreted stars,
assembled as the satellite galaxies fall into the potential well of the
main galaxies and then are disrupted. Different mechanisms take
  place during the assembly of the stellar haloes and the mass of the
  accreted satellite galaxies as well as their time of accretion play
  a major role.  In lower-mass satellites, the star formation is
  truncated earlier due to the gas exhaustion, gas outflows driven by
  Supernova, tidal stripping and/or reionization. Such quenching
  likely occurs before these clumps merged with the main galaxy, thus
  these satellites  possess mainly very old stars. On the contrary,
  more-massive satellites experience a more prolonged star formation
  activity due to their efficiency in retaining gas in the deeper
  potential wells.  These massive satellites have both young and old
  populations (Tissera et al. 2014). A combination accretion time and mass of the satellites will set the age profile. A
  negative age profile could arise when low mass satellites are
  accreted very early on.  These will contribute mostly to the inner
  regions with their old stars (Tissera et al. 2018). Intermediate
  and high-mass  satellites accreted later-on will have
  younger stellar populations (since they continued forming stars for longer
  periods) and their stars will populate
  both the inner and outer regions. The presence of a larger fraction of oldest
  early-on accreted stars in the center of the galaxy (small radii) will set the negative age
  gradient, but the strenght of such profiles depends on the
  particular assembly history of each galaxy.
 The slope is also affected by the
generally flat age profile of the \it in situ \rm stars and their
dominance in the inner-halo region. The in situ component comprises a combination of well-mixed stellar populations primarily located in the inner region of the stellar haloes. Disk-heated stars populate the inner regions of haloes increasing the fraction of younger stellar populations and contribute to
  flatten the  age profile. In some cases, like in Aq-D, a
flat age gradient can be generated by the opposite age dependance of the
\it in situ \rm and accreted stellar populations.



The Aquarius stellar halos have greater masses than expected from
observations \citep{harmsen2017}. Since the stellar-halo profiles
are consistent with an Einasto profile, most of the mass is concentrated
in the central regions \citep{tissera2014}. Part of the mass excess
could be due to inefficient regulation of the star-formation
activity or an overproduction of disk-heated stars. In the inner
halo, the fraction of disk-heated stars can differ from one halo to
another, as reported in Table 1 of \citet{tissera2013} (in
percentages: 31 (Aq-A), 3 (Aq-B), 24 (Aq-C), 26 (Aq-D), and 35 (Aq-G)).
These stars could also originate through a misclassification of
thick-disk stars or by the presence of endo-debris stars that
contribute significantly (from 20\% to 40\%), as given in Table 1 of
\citet{tissera2013}. The contribution of  endo-debris stars can be diminished by
improving the efficiency of the SN feedback.
 If the \it in situ \rm contributions were removed, or
 diminished significantly, then the stellar mass of these halos
would be more in agreement with current MW observations, and the stellar age
profiles of some of them would likely be closer to the 
 reported values ( see \citealt{monachesi2018} for similar
   conclusions using the Auriga simulations.)

In order to close the interpretation of the age profiles, we make 
use of the analysis presented in \citet{tissera2014}, where the 
mass contributions of satellites with different dynamical masses is 
investigated. From Fig. 6 in that paper, it is clear that Aq-A and 
Aq-C formed their stellar halos with important contributions from 
 less-massive satellites, while the remaining halos accreted stars 
from satellites more massive than $10^{9.5}$~M$_\odot$\footnote{
The dynamical masses (dark matter and baryons) of the satellites
galaxies are estimated before they enter the virial radius of the
progenitor using the SUBFIND algorithm}. This is 
particularly relevant for the properties of the central regions. 
Indeed, Aq-A and Aq-C show central age distributions which 
resemble the ACS observed in the MW.  Hence, the analysis of these 
simulations suggests that the MW should have formed its central 
regions by the accretion of less-massive satellites  ($< \sim
10^{9.5}$~M$_\odot$) and did not 
have a significant major-merger contribution. This is consistent 
with previous works where other methods and models have been 
adopted \citep{deason2017,dsouza2017,monachesi2018}.

\section{Summary and Conclusions}

In this paper, we focused on  an analysis of the age gradients in the 
stellar halos of a subset of MW-mass galaxies drawn from the Aquarius 
Project. We found that these stellar halos exhibit a diversity of age 
profiles, reflecting their different histories of formation and assembly. 
Our results suggest that negative age gradients are determined principally by the 
accreted component, and that the \it in situ \rm stars affect the slopes 
in the inner regions, depending upon their relative importance.
  The flatter age profile of the in-situ component is caused by the
  presence of different well-mixed stellar populations, including the
  contribution of disk heated stars and those formed by the gas
  transported inward by more massive satellites. The negative age
  gradient set by the accreted stellar component reflects the
  inside out assembly of the halo with the contribution of the latest
  merger events to the outkirts.  The characteristics of the accreted
  satellites such as mass, gas fraction, and their accretion time contribute to modulate the
  slope, making it less negative if they are able to reach the inner
  regions carrying younger stars and gas to fuel  star
  formation activity. In 
general, the \it in situ \rm component flattens the gradients of the 
global profiles. The two halos that show the steepest age gradients are 
those which formed primarily from the contributions of small- and 
 intermediate-mass satellite galaxies. Halos assembled with 
significant contributions from more-massive satellites tend to have 
shallower age gradients, because these systems carried in younger
stars and gas to feed new star formation activity. Our analysis shows 
that similar slopes of the age gradient reported in the MW's halo can be 
reproduced by considering {\it only} the contribution of the accreted 
stars. This {suggests} that the simulated galaxies {might be } producing an excess of \it 
in situ \rm stars, and that the subgrid physics should be improved to 
reduce such an overproduction of stars. The strong negative age gradient 
observed in the MW's halo is found in the simulated halos with 
important contributions from less-massive satellites, suggesting that 
the MW's halo assembled from satellites with total dynamical mass lower 
than $10^{9.5}M_{\sun}$ and a minimal contribution from {\it in situ} stars.

\acknowledgments 
PBT acknowledges partial support from Fondecyt Regular 1150334 
(CONICYT) and  UNAB Project 667/2015. TCB and DG acknowledges 
partial support for this work from Grant PHY 14-30152: The Physics 
Frontier Center/JINA Center for the Evolution of the Elements 
(JINA-CEE), awarded by the US National Science Foundation. BKG
acknowledges the support of STFC through the University of Hull
Consolidated Grant ST/R000840/1, and access to \it viper\rm, the
University of Hull High Performance Computing Facility.

\vspace{1cm}


\begin{figure}[!ht]
\begin{center}
\includegraphics[scale=0.70]{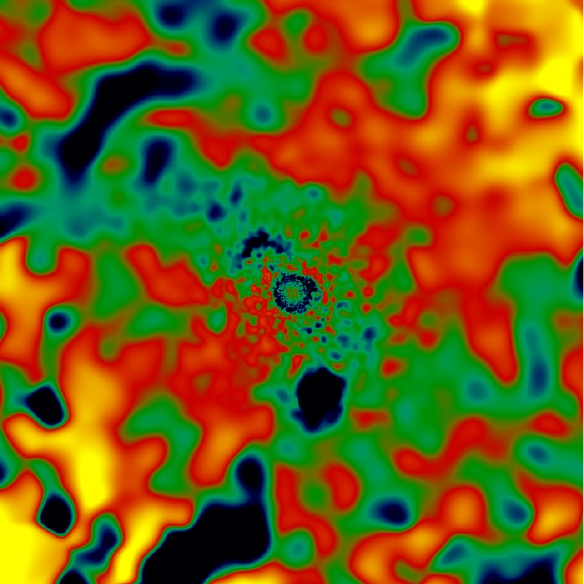}
\includegraphics[scale=0.70]{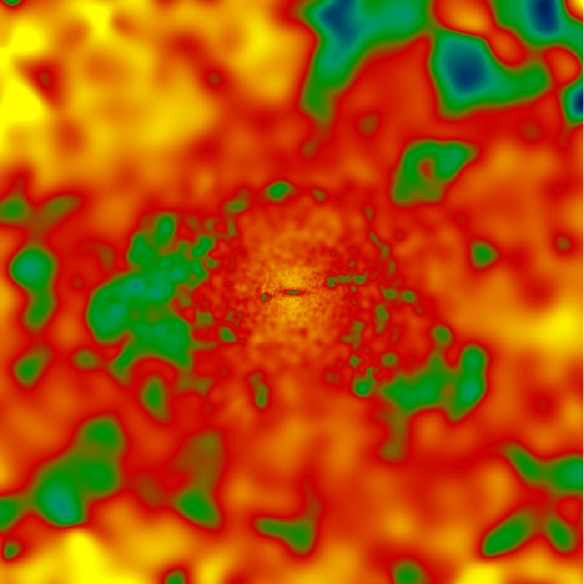}
\includegraphics[scale=0.70]{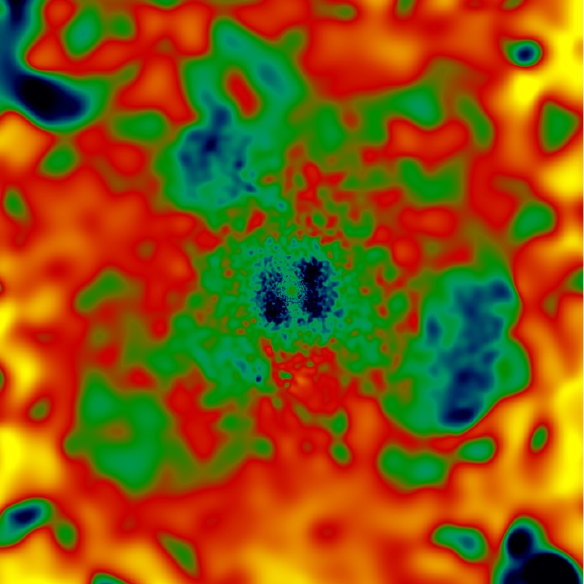}\\
\includegraphics[scale=0.70]{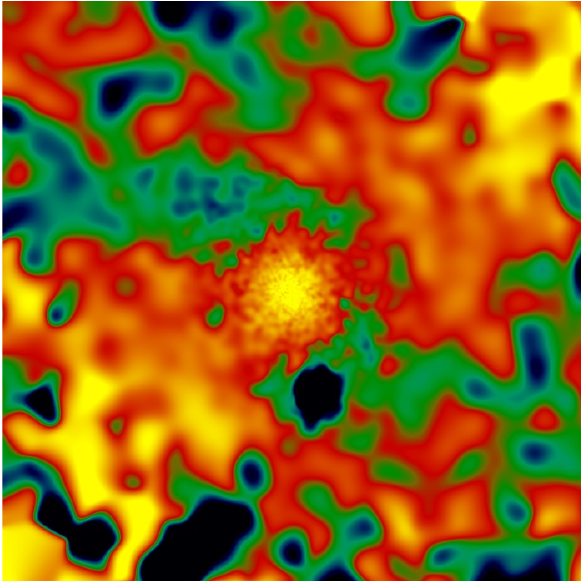}
\includegraphics[scale=0.70]{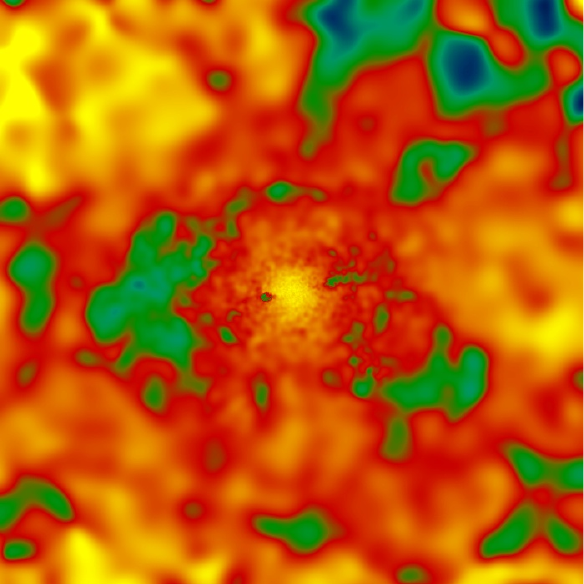}
\includegraphics[scale=0.70]{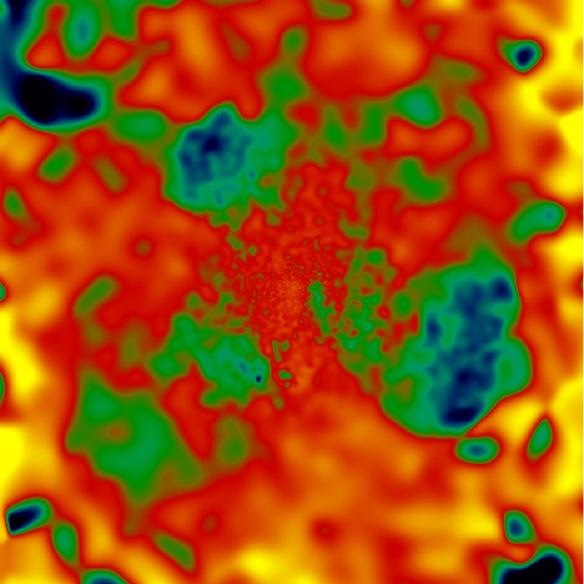}\\
\includegraphics[angle=-90,scale=1.0]{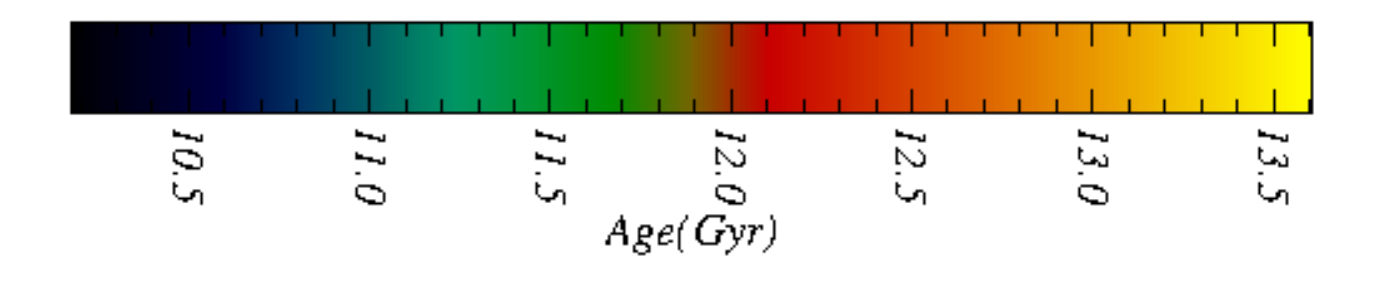}
\end{center}
\caption{Projected smoothed stellar-age maps onto the x-z plane,
  where z is the direction of rotation of the central galaxy, in the Aq-A-5 (left panel), Aq-C (middle panel) and Aq-D (right
panel) simulations. The upper panels show the age maps estimated by
considering the total stellar halo populations, while the lower panels
show the corresponding maps for the accreted stars only. The age-maps extend to 100 kpc.} 
\end{figure}

\clearpage

\begin{figure}[!ht]
\begin{center}
\includegraphics[scale=0.80]{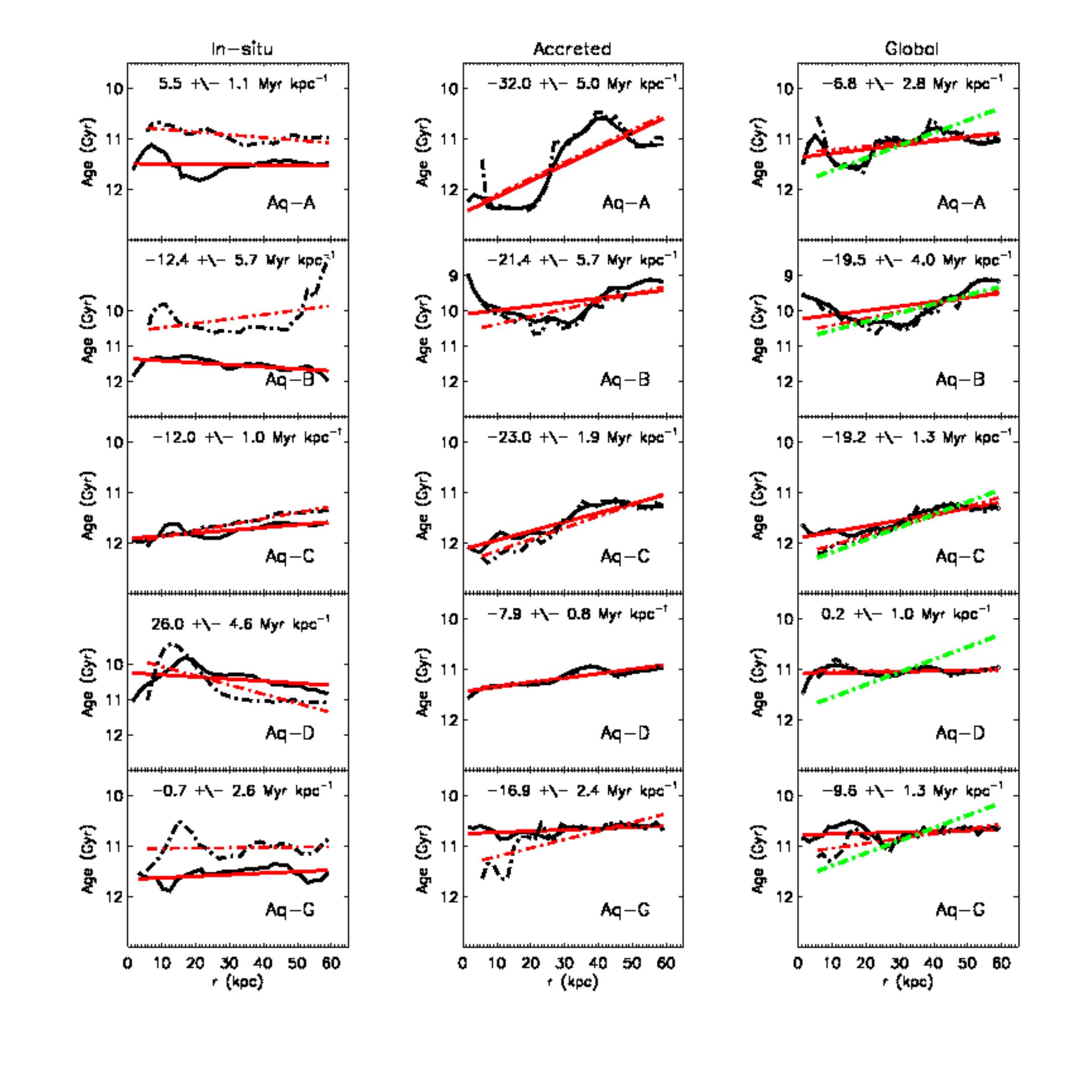}
\end{center}
\caption{The median age profiles for the \it in situ \rm (left panels), the
accreted (middle panels), and the total (right panels) stellar
populations in the five selected Aquarius halos (from top to botton:
Aq-A, Aq-B, Aq-C, Aq-D and Aq-G). The linear regressions applied to the
median age profiles are shown in all plots (red lines). The continuous
lines represent the halo samples where the spherical averages have been
implemented, while the dot-dashed lines denote the fiducial
  stellar  halo , where stellar particles within $|z| <$ 5 kpc are
  excluded to improve the comparison with observations. In all panels, the reported age gradients refer to the latter sample. The green dot-dashed lines in the right panels represent the observed age gradient in the MW.} 
\end{figure}

\clearpage

\begin{figure}[!ht]
\begin{center}
\includegraphics[scale=0.75]{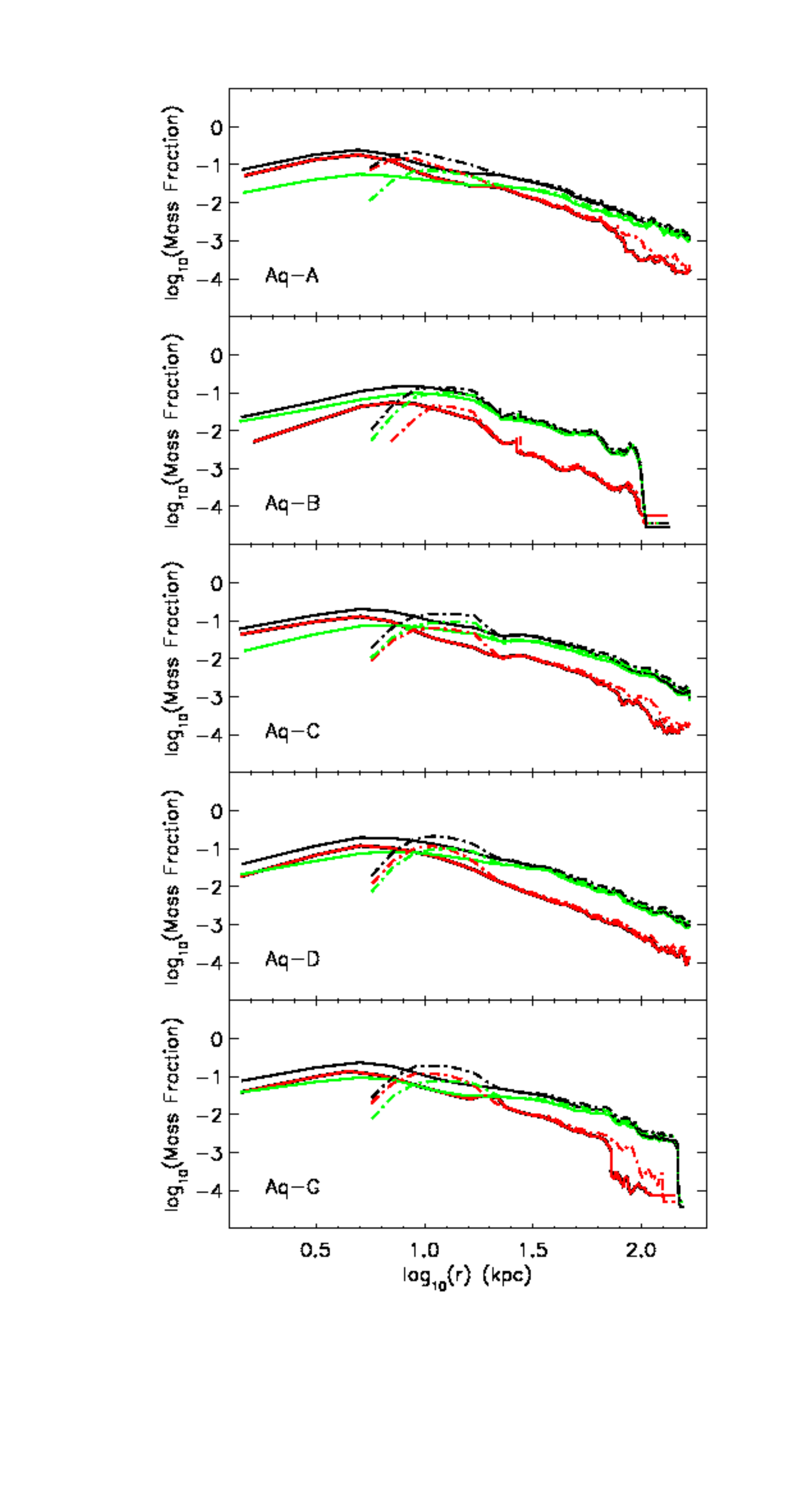}
\end{center}
\vspace{-2cm}
\caption{Mass fractions of stars formed \it in situ \rm (red lines), accreted
(green lines), and the total (black lines) stellar populations, as a
function of the galactocentric radius, for the five analyzed
halos. The mass fractions are calculated with respect to the total
stellar halo mass within the virial radius of each galaxy. Continuous and dot-dashed lines have the same meaning as in Figure 2.
The x-axis and y-axis are on logarithmic scales.} 
\label{rv}
\end{figure}

\clearpage


\end{document}